\documentstyle[prd,aps,floats]{revtex}    
\begin{document}    
\draft    
    
%
\twocolumn[\hsize\textwidth\columnwidth\hsize\csname     
@twocolumnfalse\endcsname    
    
\title{Generalized Assisted Inflation}     
\author{E.J. Copeland $^\ast $, Anupam Mazumdar $^\dagger $, and     
N.J. Nunes $^\ast$}    
\address{$^\ast $ Centre for Theoretical Physics, University of Sussex, Falmer,    
 Brighton BN1 9QJ,~~~U.~K. \\    
$^\dagger $ Astrophysics Group, Blackett Laboratory, Imperial College,     
London, SW7 2BZ,~~~U.~K.    
}      
\date{\today}     
\maketitle    
\begin{abstract}    
We obtain a new class of exact cosmological solutions for    
multi-scalar fields with exponential potentials. We    
generalize the assisted inflation solutions previously obtained,     
and demonstrate how they are modified when there exist cross-couplings between  
the fields, such as occur in supergravity inspired cosmological models.

\end{abstract}

\pacs{PACS numbers: 98.80.Cq \hspace*{1.5cm} IMPERIAL-AST     
99/4-4, astro-ph/yymmmnn }    
\vskip2pc]    
    
    
\section{Introduction}    
Scalar field theory has become the generic playground for building    
cosmological models related to particle physics, in particular for    
obtaining inflationary cosmologies. One such class of models involve    
exponential potentials (exp$(\sqrt{16 \pi/p~m_{\rm Pl}^2} \phi)$), which  
lead to power law inflation, $a \propto t^p$, with $p>1$, for sufficiently    
flat potentials \cite{lucchin,halliwell,burd}. A number of related features have  
also been discovered for    
such potentials: in a Universe containing a perfect fluid and such a scalar  
field, then for a wide range of    
parameters    
the scalar field `mimics' the perfect fluid, adopting its equation    
of state \cite{wetterich,wands93,joyce} and leading to attractor scaling    
solutions at late time \cite{copeland98}. These solutions     
offer a plausible mechanism for stabilizing the dilaton field in    
models of gaugino condensation arising in supersymmetry breaking  
\cite{barreiro}.    
   
It is generally assumed that even if there are many scalar fields present,    
only one of them will dominate the dynamics and roll down the    
potential slowly. However, recently Liddle, Mazumdar and Schunck  
(LMS)\cite{lms}, have demonstrated in a    
particular    
example that multiple scalar fields, each with an exponential potential,    
can lead to inflationary solutions,      
even if the individual field potentials are too steep for inflation.    
There exists a cumulative effect of all the fields that can give rise to   
inflationary behavior - a result    
they termed as ``Assisted Inflation''.    
Malik and Wands have demonstrated that the associated attractor solution could  
be identified through a rotation    
in field space, with a hybrid model where the vacuum energy  had an exponential  
dependence upon the dilaton    
field \cite{malik}.   
Multiple exponential potentials do arise in modern Kaluza-Klein    
theories. Indeed they are a natural outcome of the compactification of    
higher dimensional theories down to 3+1 dimensions. With this in mind it    
is worth investigating such potentials in a bit more detail. Indeed, Kanti and  
Olive have recently proposed a possible realisation of assisted inflation based  
on the compactification of a five-dimensional Kaluza Klein model \cite{kanti}.  
It also raises the question,   
could inflation arise out of the 11 dimensional supergravity models    
compactified on squashed seven spheres for example?    
Such models have been investigated by a number of authors as    
low energy cosmologies from string or M-theory    
\cite{lukas,stelle,reall,reall1}.     
Most `realistic' models of dimensional reduction lead to steep potentials,   
which generally do not lead to    
inflation. In this paper we consider a more    
general class of exponential potentials, which can include those generally    
found in supergravity compactifications, and obtain exact cosmological solutions    
for them. In particular we demonstrate how difficult it is to obtain   
assisted inflation when there exists cross-couplings between the scalar fields  
in the potential, a result also discussed in \cite{reall} and \cite{kanti}. We  
first recall the    
model discussed by LMS,     
before generalizing their potential to exponentials involving cross-coupling  
terms and demonstrating that the attractor behavior of the scalar-fields still    
exists, leading to   
scaling  solutions for the generalized potential. We then turn our    
attention to the case of potentials involving multiple exponential terms  
containing the same scalar fields, and relate the solutions to those arising    
in supergravity models.

    
\section{The dynamics of assisted inflation}    
   
Liddle et al\cite{lms}, considered  $n$ scalar fields, $\phi_i ,\,i=1..n$, each    
with exponential potentials decoupled from each other:    
\begin{equation}    
V_i(\phi_i)  =   V_0 \exp \left( \alpha_i \, \phi_i \right) \,,     
\end{equation}    
where $\alpha_{i}$ is the slope of the individual field with    
dimensions of inverse Planck mass. Although the fields are not directly    
coupled through the potential, they are coupled through the Friedmann equation,    
which implies that the combined role of the     
fields affect the expansion rate of the Universe:    
\begin{eqnarray}    
\label{Motion}    
H^2 & = & \frac{8\pi}{3m_{{\rm Pl}}^2} \sum_{i=1}^n \left[ V_i(\phi_i) +     
\frac{1}{2} \dot{\phi}_i^2 \right] \,, \\    
\label{eqofmo}    
\ddot{\phi}_i & = & - 3 H \dot{\phi}_i - \frac{dV_i(\phi_i)}{d\phi_i} \,,    
\end{eqnarray}    
where $H=\dot{a}/a$ is Hubble's constant and $a$ is the scale factor of the    
flat FRW Universe. The solution to this is the modified power law \cite{lms}    
\begin{equation}    
\label{power}    
a(t)  \propto  t^p \,,    
\end{equation}    
where $p$ is given by    
\begin{equation}    
\label{slope}    
p = \frac{16 \pi}{m_{{\rm Pl}}^2} \sum_{i=1}^n \frac{1}{\alpha_i^2} \,.    
\end{equation}    
Inflationary solutions exist provided $p > 1$, hence even if each of the  
$\alpha_i$'s are too steep to    
individually satisfy the condition    
for inflation, as long as $n$ is large enough, the inequality $p>1$    
can be satisfied. These solutions, and the inflationary ones we shall present  
below are eternal, they do not possess an exit from the  
inflationary epoch. Realistic models would of course have to possess 
such an exit in  
order to enter the radiation and matter dominated epochs of our Universe. 
The particular example of Eq.~(\ref{slope}) suggests that it is    
worth investigating whether or not such assisted inflation exists with more    
general exponential potentials. An alternative approach with interesting results  
has been adopted in \cite{kanti}, where they have applied the assistance method  
to the case of polynomial scalar potentials.    
\subsection*{Exponential potentials with coupled scalar fields.}   
 
To begin with we consider the most natural generalization of    
the single field exponential, namely the case of two coupled fields:   
\begin{equation}    
\label{trial}    
V(\phi,\psi) = V_0 e^{\alpha \phi + \beta \psi} \,,    
\end{equation}    
where $\alpha$ and $\beta$ are the slopes for the fields $\phi$    
and $\psi$.    
we see from Eq.(\ref{Motion})     
that dimensionally, the right hand side should decrease    
as $t^{-2}$, because $H^2 \propto t^{-2} \propto V_i(\phi_{i})$.    
We further     
assume that for our potential Eq.(\ref{trial}),    
\begin{equation}    
\label{assum1}    
e^{\alpha \phi} = \frac{k_{\alpha}}{t^c}, ~~~~    
e^{\beta \psi} = \frac{k_{\beta}}{t^{2-c}} \,,    
\end{equation}    
where $k_{\alpha}$, $k_{\beta}$ are dimensional constants and $c$ is a    
dimensionless constant. Substituting the power law solution Eq.(\ref{power}) 
into  Eq.(\ref{Motion}) we obtain:   
\begin{eqnarray}    
\label{evolv}    
p^2 & = & H^2 t^2 \, \nonumber \\     
 & = & \frac{8\pi}{3m_{{\rm Pl}}^2} \left(V_0 k_{\alpha}k_{\beta} +     
\frac{1}{2} \left(\frac{c}{\alpha}\right)^2 +     
\frac{1}{2} \left(\frac{2-c}{\beta}\right)^2 \right) \,,    
\end{eqnarray}    
which when coupled with Eq.(\ref{eqofmo}) for    
the $\phi$ and $\psi$ fields leads to,   
\begin{equation}    
\label{sol1}    
V_0 k_{\alpha}k_{\beta} = \frac{(3p-1)c}{\alpha^2}  \,, ~~~    
V_0 k_{\alpha}k_{\beta} = \frac{(3p-1)(2-c)}{\beta^2} \,,    
\end{equation}    
hence   
\begin{equation}    
\label{sol3}    
V_0 k_{\alpha}k_{\beta} = \frac{2(3p-1)}{\alpha^2+\beta^2} \,.    
\end{equation}    
Using Eq.(\ref{sol1}--\ref{sol3}) in Eq.~(\ref{evolv}), we obtain    
a simple scaling solution between the two fields:    
\begin{eqnarray}    
\label{scal}    
p = \frac{16\pi}{m_{{\rm Pl}}^2} \frac{1}{\alpha^2+\beta^2} \,, \\    
\left(\frac{\dot{\phi}}{\dot{\psi}} \right)^2 =     
\left(\frac{\alpha}{\beta} \right)^2.     
\end{eqnarray}    
An important problematic feature for inflationary solutions   
confronts us immediately in Eq.~(\ref{scal}), namely the coupling between the   
two fields reduces the rate of expansion of the Universe, a point also made in   
\cite{reall,kanti}. We will return to this point again later. An alternative 
method which would also lead to Eq.~(\ref{scal}) is described in 
\cite{malik} in terms of field rotations, which results in the introduction 
of two orthogonal fields, one of which is massless and the 
other posseses an exponential potential. 
  
Having demonstrated that it is possible to obtain a scaling solution without    
using slow roll approximations, we now generalize this simple     
case to include an arbitrary number of fields and exponential terms    
making up the overall potential.

\subsection*{General exponential potentials with coupled scalar fields.}   
   
We now consider a potential where we have multiple scalar fields but their     
corresponding exponential potentials can contain arbitrary combinations of    
the fields with different slopes. The potential we will consider is:    
\begin{equation}    
\label{pot}    
V =  \sum_{s=1}^n V_s = V_0 \sum_{s=1}^n \exp \left(\sum_{j=1}^{m_s} \alpha_{sj}  
\,    
 \phi_{sj} \right) \,, \\    
\end{equation}    
with the corresponding Friedmann equation :    
\begin{equation}    
\label{genevol0}    
H^2  =  \frac{8\pi}{3m_{{\rm Pl}}^2}  \left[ V +     
\sum_{s=1}^n \sum_{j=1}^{m_s} \frac{1}{2} \dot{\phi}_{sj}^2 \right] \,,    
\end{equation}    
where, from now on $q,r,s$ stands for index terms in the potential and $i,j,k,l$  
stands for field indexes, hence, $\phi_{sj}$ stands for the $j$th field in the  
$s$th potential term. In other words, there are a total of $\sum_{s=1}^n m_s$  
fields distributed in groups of $m_s$ through the terms of the potential. We  
obtain the solution to this problem by generalising the assumption  
Eq.(\ref{assum1}). There exists an  
attractor region with a power law solution, which from  
Eq.~(\ref{genevol0}), dimensionally satisfies $H^2 \propto t^{-2} \propto V_i$.  
Hence we write,  
\begin{eqnarray}   
\label{genevol}   
e^{\alpha_{sj} \phi_{sj}} & = & \frac{k_{sj}}{t^{c_{sj}}} \,, \\    
\label{csj}  
\sum_{j=1}^{m_s} c_{sj} & = & 2 \,,    
\end{eqnarray}    
where $k_{sj}$ are dimensional and $c_{sj}$ are dimensionless constants    
respectively.    
Eq.(\ref{genevol}), coupled with the equations of motion:    
\begin{equation}    
\label{gensol1}    
\ddot{\phi}_{sj} + 3 H \dot{\phi}_{sj} + \frac{\partial  V}    
{\partial \phi_{sj}} = 0 \,,    
\end{equation}   
result in:     
\begin{equation}    
(3p-1) c_{sj} = \alpha_{sj}^2 V_0 \prod_{k=1}^{m_s} k_{sk} \,,    
\end{equation}    
from which we find, using Eq.(\ref{csj}) and Eq.(\ref{gensol1}):   
\begin{eqnarray}     
\label{gensol2}    
V_0 \prod_{k=1}^{m_s} k_{sk} & = & \frac{2 (3p-1)}{\sum_{j=1}^{m_s}    
 \alpha_{sj}^2} \,, \nonumber \\   
\left(\frac{c_{sj}}{\alpha_{sj}} \right)^2 & = & \frac{4 \alpha_{sj}^2}    
{\left(\sum_{k=1}^{m_s} \alpha_{sk}^2 \right)^2} \,.     
\end{eqnarray}    
When substituted into Eq.(\ref{genevol0}) this leads to a key result,    
the generalisation of the original assisted inflation result given by   
Eq.~(\ref{slope}):    
\begin{equation}   
\label{slope1}    
p = \frac{16 \pi}{m_{{\rm Pl}}^2} \sum_{s=1}^n \frac{1}{\sum_{j=1}^{m_s}    
\alpha_{sj}^2} \,.    
\end{equation}   
We also note that the generalization of the scaling solution found in  
Eq.(\ref{scal}), quickly follows for    
the case of any two scalar fields, $\phi_{sj}$ and $\phi_{ql}$:     
\begin{equation}    
\label{scal1}    
\left(\frac{\dot{\phi}_{sj}}{\dot{\phi}_{ql}} \right)^2 =     
\left(\frac{\alpha_{sj}}{\alpha_{ql}}\right)^2     
\left(\frac{\sum_{i=1}^{m_{q}} \alpha_{qi}^2}{\sum_{k=1}^{m_s}     
\alpha_{sk}^2}\right)^2  \,.    
\end{equation}     
It is directly evident that Eqs.(\ref{slope1}) and (\ref{scal1})    
reduce to Eq.~(\ref{slope}) for the   
example of $n$ exponential terms each containing just one field, and   
Eq.~(\ref{scal}) for the case of one exponential term but containing two fields.  
We again see the inhibiting affect that multiplicative coupling of the fields  
(i.e. $m_s>1$) has for obtaining inflationary solutions. However, in this case,  
as with the original version of assisted inflation, this can be compensated for  
if there are enough exponential terms present in the potential (i.e. if $n$ is  
large enough)\cite{lms}.  
 
There is another feature of the potentials we have been discussing so    
far that makes them rather unphysical    
in general.    
We have been demanding that any two fields present can not be the same,   
(i.e. $\phi_{sj} \neq \phi_{ql}$). In other words they can only appear once in  
the full potential.    
Nearly all realistic models which emerge from compactifications arising    
in supergravity models have the same field appearing in at least two    
separate exponential terms. In the following section, we turn our attention    
to this case.        
\section{Exponential potentials inspired by supergravity models.}    
  
To set the scene, we generalise Eq.~(\ref{trial}) to the case where the scalar  
field potential takes the following form:     
\begin{equation}   
\label{2x2} 
V(\phi_1,\phi_2) = z_1 e^{\alpha_{11} \phi_1+ \alpha_{12} \phi_2} + z_2  
e^{\alpha_{21} \phi_1 + \alpha_{22} \phi_2}  
\end{equation}    
where, $\alpha_{sj} $, are dimensional constants which can take any real value  
and $z_s>0$. The occurrence of 
such forms of the potential are quite common in dimensionally reduced   
supergravity models \cite{stelle,reall}. Remarkably we can solve this system to  
obtain scaling solutions in a manner analogous to those already presented in  
Eq.~(\ref{assum1})-(\ref{scal})  obtaining the unique late time scaling solution  
for the fields $\phi_1$ and $\phi_2$,   
\begin{equation}   
\label{scalnew}   
p = {16\pi \over m_{{\rm Pl}}^2} \left[{(\alpha_{21} - \alpha_{11})^2 +  
(\alpha_{22} - \alpha_{12})^2 \over (\alpha_{11} \alpha_{22} - \alpha_{21}  
\alpha_{12})^2 } \right].  
\end{equation} 
This simple result reduces to the particular cases we have already investigated  
when the appropriate limits are taken. For example, when  
$\alpha_{21}=\alpha_{12}=0$, we reproduce the result Eq.~(\ref{slope}). The  
equivalent of the assisted inflation result Eq.~(\ref{scal}) follows by setting  
$\alpha_{11} \alpha_{21} + \alpha_{12} \alpha_{22} =0$, in which case we find,  
\begin{equation}  
\label{scalnew1}    
p = {16\pi \over m_{{\rm Pl}}^2} \left[{1 \over \alpha_{11}^2 + \alpha_{12}^2} +  
{1 \over \alpha_{21}^2 + \alpha_{22}^2} \right].     
\end{equation} 
 
We shall now generalize the potential to $n$ such exponential potentials   
and $m$ combinations of linear fields in the exponent (explicitly calculating  
for the simple case of 2 terms $\times$ 2 fields of Eq.~(\ref{2x2})). The  
generalised Eq.~(\ref{2x2}) is then 
\begin{equation}   
V = \sum_{s=1}^n z_s \exp \left( \sum_{j=1}^{m} \alpha_{sj} \phi_j \right)  \,.   
\end{equation}    
Of course we are allowing here for the possibility that $\alpha_{sj} =0$ for  
some combination of $sj$.  
We assume that for late times the fields have an attractor solution, given by    
\begin{equation} 
\label{expon} 
z_s\exp \left( \sum_{j=1}^m \alpha_{sj} \phi_j \right) = \frac{k_s}{t^2} \,, 
\end{equation} 
and, following Eq.~(\ref{genevol}) and Eq.~(\ref{csj}) we write  
\begin{equation}   
\label{late}   
\phi_j= a_j - \frac{c_{sj}}{\alpha_{sj}} \ln t \,,   
\end{equation}   
where, $a_j$ is a constant depending on the initial conditions and $\sum_{j=1}^m  
c_{sj} =2, ~~~s=1..n$. Substituting Eq.(\ref{late}) into the equation of motion  
Eq.(\ref{eqofmo}), we obtain the constraint equation for the $j$th field, which  
follows from assuming the existence of a power law solution, 
\begin{equation}  
\label{cijgeneral}  
(3p-1) \frac{c_{sj}}{\alpha_{sj}} = \sum_{q=1}^n \alpha_{qj}k_q \,.   
\end{equation}   
Again, using $\sum_{j=1}^m c_{ij} =2$ we obtain,  
\begin{equation} 
\sum_{j=1}^m \alpha_{sj} \left( \sum_{q=1}^n \alpha_{qj} k_q \right) =  
2(3p-1) \,, 
\end{equation}    
which is equivalent to writing 
\begin{equation} 
\label{seqn} 
\sum_{q=1}^n A_{sq} k_q = 2(3p-1) \,, 
\end{equation} 
where  
\begin{equation} 
\label{defa} 
A_{sq} = \sum_{j=1}^m \alpha_{sj} \alpha_{qj} \,. 
\end{equation} 
Since $s$ is a free index, we have a set of $n$ equations that can be witten as 
\begin{equation} 
\label{eqk} 
A {\bf k} = 2(3p-1) \,, 
\end{equation} 
where $A$ is the $n \times n$ matrix with elements $A_{sq}$ 
and ${\bf k}=(k_1,..,k_n)$ a column vector. For the 2 $\times$ 2 case of  
Eq.~(\ref{2x2}) we obtain 
\begin{equation} 
A = \left( \begin{array}{cc} \alpha_{11}^2+\alpha_{12}^2 &  
\alpha_{11}\alpha_{21}+\alpha_{12}\alpha_{22}  \\  
\alpha_{21}\alpha_{11}+\alpha_{22}\alpha_{12}  & \alpha_{21}^2+\alpha_{22}^2  
\end{array} \right) \,. 
\end{equation} 
The solution to this system is 
\begin{equation} 
{\bf k} = A^{-1} 2(3p-1) \,, 
\end{equation} 
with 
\begin{equation} 
\label{defb} 
A^{-1} = \frac{A_{COF}^T}{det A}  \,, 
\end{equation} 
where $A_{COF}^T$ is the transpose of the cofactor matrix of $A$. 
To simplify notation we will write $B \equiv A_{COF}^T$ and the sum of the  
elements in row $s$ of $B$ as $B^s \equiv \sum_{q=1}^n B_{sq}$, hence, each  
$k_s$ is  
\begin{equation} 
\label{defks} 
k_s = \frac{2(3p-1)}{det A} B^s \,. 
\end{equation} 
For the 2 $\times$ 2 case this yields 
\begin{eqnarray} 
\label{bkk} 
B &=& \left( \begin{array}{cc} \alpha_{21}^2+\alpha_{22}^2 &  
-\alpha_{11}\alpha_{21}-\alpha_{12}\alpha_{22}  \\  
-\alpha_{21}\alpha_{11}-\alpha_{22}\alpha_{12}  & \alpha_{11}^2+\alpha_{12}^2  
\end{array} \right) \,, \\ 
k_1 &=& 2(3p-1) \frac{\alpha_{21}^2+\alpha_{22}^2  
-\alpha_{11}\alpha_{21}-\alpha_{12}\alpha_{22}}{(\alpha_{11}\alpha_{22}-\alpha_{ 
12}\alpha_{21})^2} \,, \\ 
k_2 &=& 2(3p-1)  
\frac{-\alpha_{21}\alpha_{11}-\alpha_{22}\alpha_{12}+\alpha_{11}^2+\alpha_{12}^2 
}{(\alpha_{11}\alpha_{22}-\alpha_{12}\alpha_{21})^2} \,. 
\end{eqnarray} 
 
>From Eqs.~(\ref{late}) and (\ref{cijgeneral}) the late time ratio between the  
kinetic terms of two different fields becomes 
\begin{equation}    
\left( \frac{\dot{\phi}_j}{\dot{\phi}_l} \right)^2 = \left( \frac{\sum_{q=1}^n  
\alpha_{qj}B^q}{\sum_{r=1}^n \alpha_{rl} B^r} \right)^2 \,. 
\end{equation} 
 
Substitution of Eqs.~(\ref{defks}) and (\ref{cijgeneral}) into the Friedmann  
equation yields, 
\begin{eqnarray}     
p^2 &=& \frac{8\pi}{3m_{{\rm Pl}}^2} \left[ \sum_{s=1}^n k_s  + \frac{1}{2}  
\sum_{j=1}^m \left(\frac{c_{sj}}{\alpha_{sj}} \right)^2 \right] \nonumber \, \\   
    &=& \frac{8\pi}{3m_{{\rm Pl}}^2}\left[ 2(3p-1)\sum_{s=1}^n  
    \frac{B^s}{det A} +2\sum_{j=1}^m \left(\frac{\sum_{q=1}^n  
\alpha_{qj}B^q}{det A}\right)^2 \right] \nonumber \,.  
\end{eqnarray} 
After some algebra, we obtain the simple result for $p$ as the ratio between the  
sum of all the elements in the cofactor matrix of $A$ and its determinant. 
\begin{equation} 
\label{finalp} 
p = \frac{16\pi}{m_{{\rm Pl}}^2}\frac{\sum_{s}^n \sum_{q}^n B_{sq}}{det A} \,. 
\end{equation} 
The reader should have no problem showing that for the 2 $\times$ 2 case this  
reduces to Eq.(\ref{scalnew}). 
 
 It is instructive to rewrite Eq.~(\ref{finalp}) in another form. From  
Eqs.~(\ref{seqn}), (\ref{defks}) and (\ref{finalp}) 
\begin{equation} 
p = \frac{16\pi}{m_{{\rm Pl}}^2}\sum_{s=1}^n \frac{1}{\sum_{q=1}^n A_{sq}  
k_q/k_s} \,, 
\end{equation} 
and using Eq.~(\ref{defa}) with $q=s$ and Eq.~(\ref{defks}) we obtain, 
\begin{equation} 
\label{presult} 
p = \frac{16\pi}{m_{{\rm Pl}}^2}\sum_{s=1}^n \frac{1}{\sum_{j=1}^m \alpha_{sj}^2  
+ \sum_{q \neq s}^n A_{sq} B^q/B^s} \,. 
\end{equation} 
 
A number of points need to be made about Eq.~(\ref{presult}). It is similar in  
form to Eq.~(\ref{slope1}), which should not be too surprising, the  
additional terms in the denominator arising from the fact that we have allowed  
for fields to appear more than once in the potential, hence leading effectively  
to `self-interaction' type contributions. Indeed if these terms were turned off  
we would reproduce the result in Eq.~(\ref{slope1}). In the 2 $\times$ 2 case,  
it is the constraint leading to Eq.~(\ref{scalnew1}). Due to the presence of  
these `self-interaction' terms, $p$ could increase above the value in  
Eq.~(\ref{slope1}) if there happened to be a combination of positive and  
negative slopes in Eq.~(\ref{presult}).  
 
An issue emerges when considering these more complicated  
potentials. For the two field, two term case of Eq.(\ref{2x2}), if  
$\alpha_{11} > \alpha_{12}$ then, a necessary (but not sufficient) condition  
for the second term to be comparable to the first term at late times is,  
$\alpha_{21} < \alpha_{22}$. By comparable we mean that the potential terms  
reach a constant ratio. If this were  
not the case, then one of the two terms would quickly dominate the  
overall dynamics. 
One way to check if a combination of terms in a given potential will be  
comparable at late times is to use Eq.~(\ref{defks}) to obtain the ratios  
$k_s/k_q$ for these terms. From  
Eq.~(\ref{expon}) it follows that for consistency we require them to be  
positive, with $p>1/3$.   
In general the surviving terms (i.e. those which remain comparable) will be the  
ones with the smallest slopes,  
corresponding to the largest values of $p$. 
 
\section{Conclusions}  
In this paper we have derived a new class of exact cosmological solutions   
involving exponential potentials. In doing so we have been able to generalize  
the assisted inflation   
solutions discussed by LMS \cite{lms} to the case of multiple exponential terms  
involving many fields. Such potentials are more likely to arise in realistic  
models of particle physics where individual fields will occur in a number of  
separate exponential terms, leading to cross-couplings between the terms. In  
general, it transpires that it is more difficult to obtain assisted inflation in  
such models, the fields in any one exponential term tend to conspire to act  
against one another rather than assist each other, a result also noticed in  
\cite{reall,kanti}. This is the real reason why such models tend to fail to  
produce inflationary solutions in supergravity models compactified on squashed  
seven spheres \cite{reall}.  
We also investigated the case where a number of exponential terms contained the  
same scalar fields and demonstrated that a number of novel features emerged, 
including the possibility of increasing the rate of expansion when there 
exists a mixture of positive and negative slopes in the potential.

\acknowledgements  
We would like to thank Andrew Liddle, Karim Malik, David Wands and Orfeu  
Bertolami for useful discussions. EJC was supported   
by PPARC and is particularly grateful to David Wands for conversations on the  
nature of the generalised solutions. NJN was supported  by FCT (Portugal) under  
contract  PRAXIS XXI BD/15736/98 and AM was supported by the INLAKS and the ORS  
award.

\end{document}